\documentclass[11pt]{article}
\usepackage{moriond,epsfig}

\def\Journal#1#2#3#4{{#1} {\bf #2}, #3 (#4)}

\def\PRL{\em Phys. Rev. Lett.}
\def\PRD{{\em Phys. Rev.} D}

\def\be{\begin{equation}}
\def\ee{\end{equation}}
\def\bea{\begin{eqnarray}}
\def\eea{\end{eqnarray}}

\begin{document}
\vspace*{4cm}
\title{Jet Studies at CMS and ATLAS}

\author{ Konstantinos Kousouris}

\address{Fermi National Accelerator Laboratory, P.O. Box 500 Batavia IL 60510, USA}

\maketitle\abstracts{
The jet reconstruction and jet
energy calibration strategies adopted by the CMS and ATLAS experiments are presented.
Jet measurements that can be done
with early data to confront QCD at the highest transverse momentum scale and search
for new physics are described.
}
\section{Introduction}
Jet final states will be the dominant ones at the LHC p-p collisions. The understanding of
the jet objects will be critical for the re-discovery of the standard model and at the same time
will increase the sensitivity to new physics signals. Despite the large experimental uncertainties,
the CMS~\cite{bib:CMS} and ATLAS~\cite{bib:ATLAS} experiments will be able to probe the highest transverse momentum
scale, far beyond the Tevatron reach, even with small amount of data ($\mathcal{O}(10\,pb^{-1})$).

\section{Jet Properties}
\subsection{Jet Reconstruction}
The reconstruction of the jets can be done with different inputs (calorimeter energy depositions,
combined calorimeter and tracker information, tracks alone or particle flow candidates) and both experiments
plan to use all the above jet flavours in order to optimise the sensitivity of each physics channel. For QCD studies which
reach the highest $p_T$ and cover the full spectrum and detector acceptance, the calorimeter jets are used. The algorithms
employed in the jet reconstruction need to be infrared and collinear safe to allow theoretical calculations.
The ATLAS experiment plans to use the \textit{Seeded Cone} algorithm with two radius sizes ($R=0.4, 0.7$)~\cite{bib:ATLAS_TDR}
and the successive recombination $\mathit{k_T}$ algorithm~\cite{bib:KT} with two distance parameters ($D=0.4, 0.6$). The CMS
experiment plans to use the \textit{Seedless Cone}~\cite{bib:SISCONE} with two radius sizes ($R=0.5, 0.7$) as well as the
$\mathit{k_T}$ algorithm with two distance parameters ($D=0.4, 0.6$).

\subsection{Jet Energy Scale}
\begin{figure}[htbp]
  \begin{center}
     \includegraphics[height=4.6cm]{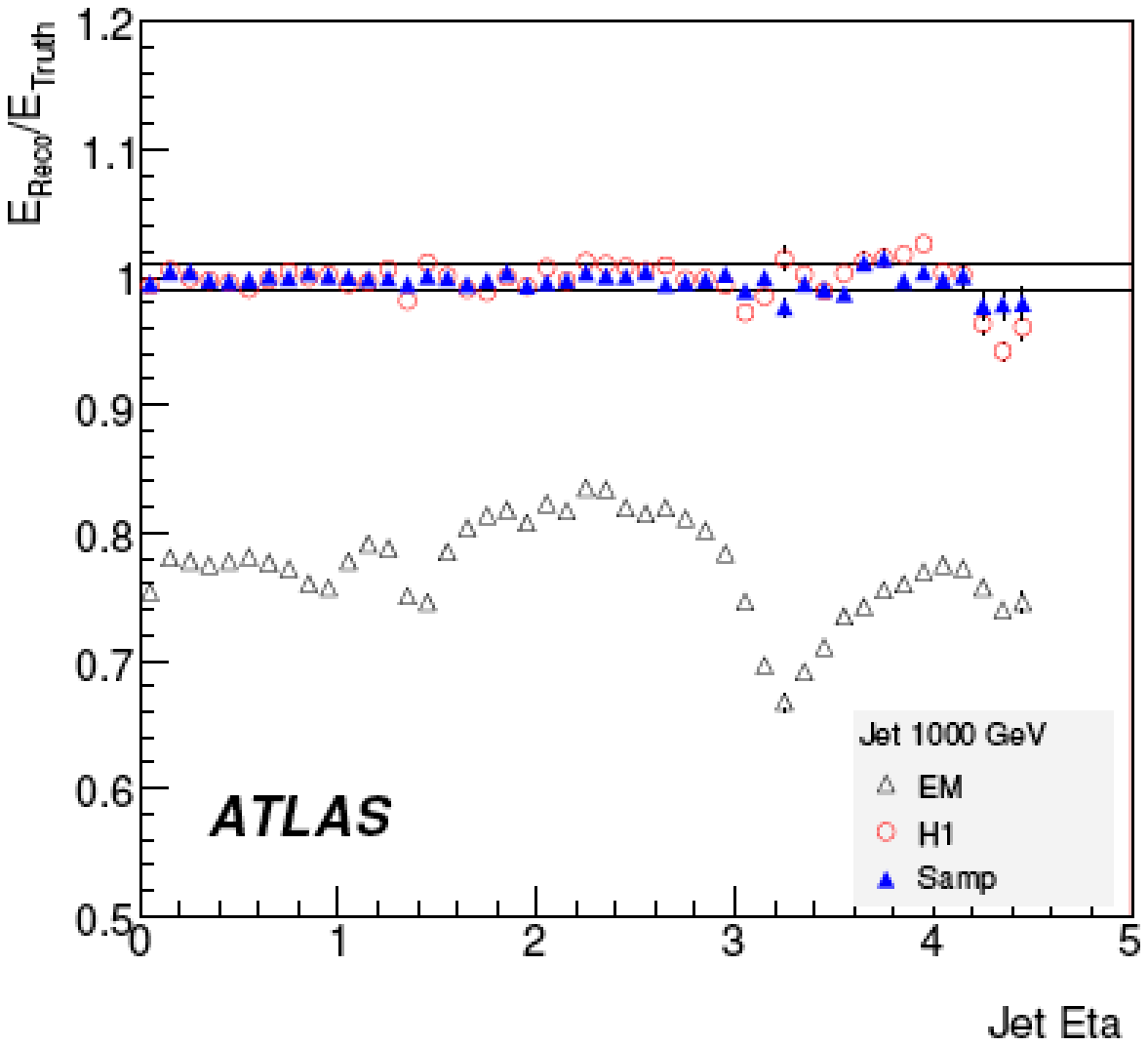}
     \includegraphics[height=4.6cm]{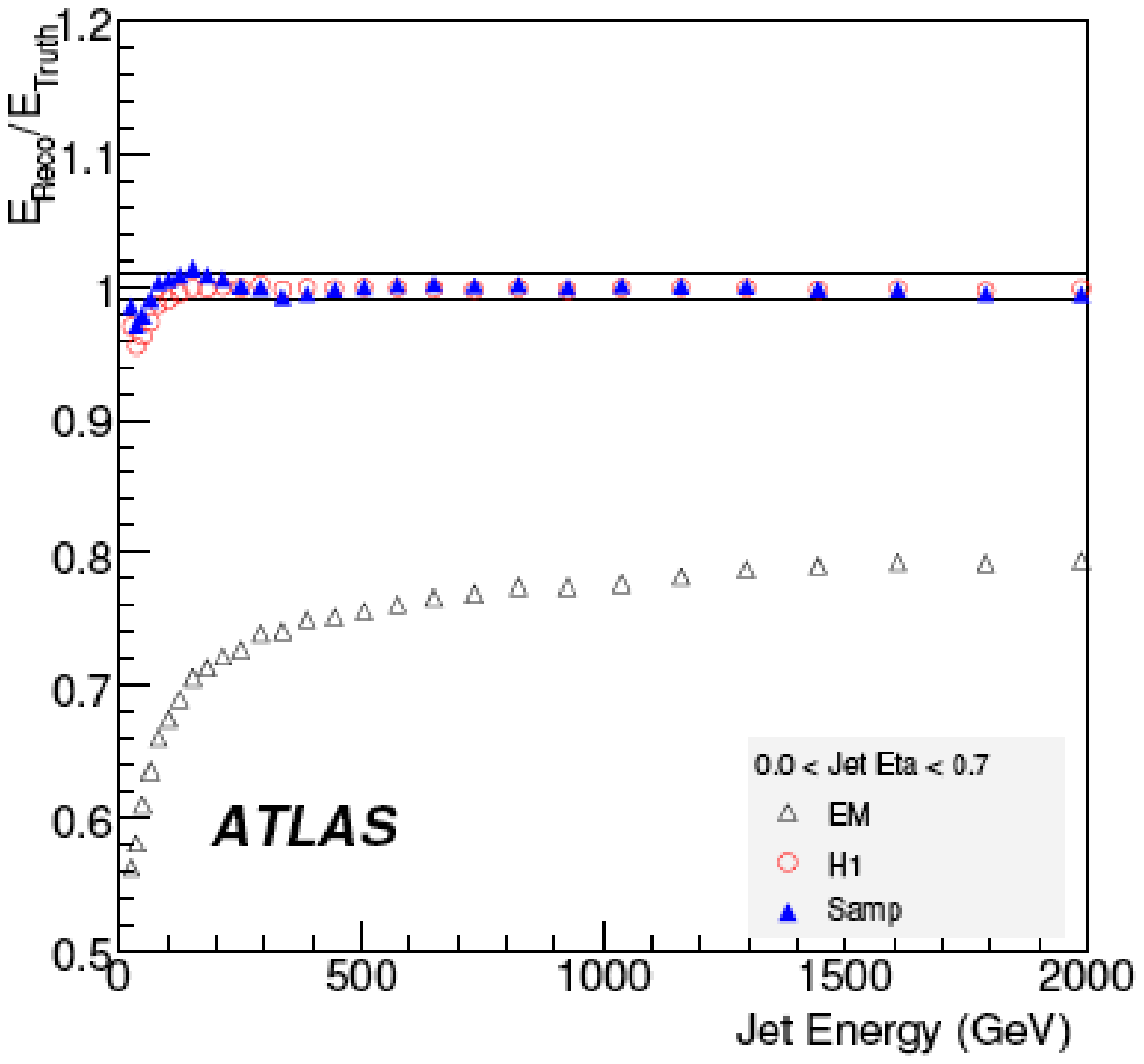}
    \caption{Effect of the jet energy corrections in the ATLAS simulation. Left: corrected jet response compared to raw jet response as a function of $\eta$. Right: corrected jet response compared to raw jet response vs energy.}
    \label{fig:jes}
  \end{center}
\end{figure}

The most important uncertainty related to jets is the jet energy scale (JES). Due to the non linear and non uniform
response of the CMS and ATLAS hadron calorimeters, it is necessary to apply jet energy corrections which
restore on average the JES. Both experiments plan to use data-driven approaches for in-situ jet calibration at startup
(dijet balancing to restore the pseudorapidity uniformity and $\gamma,Z+jet$ balancing to restore the non-linearity)
~\cite{bib:CMS_JES_PLANS}~\cite{bib:ATLAS_TDR}. At a later stage of the experiments, when Monte Carlo simulations
tuned to data will be available, the JES will be determined from the Monte Carlo. In Fig.~\ref{fig:jes} the corrected jet energy
response in the ATLAS calorimeter is shown as a function of $\eta$ and energy, compared to the raw jet response.
The systematic uncertainty of the JES expected with early data is of the order of $10\%$ and could reach $5\%$ with
$100\,pb^{-1}$ of data.

\section{Inclusive Jets}

\begin{figure}[htbp]
  \begin{center}
    \includegraphics[height=4.6cm]{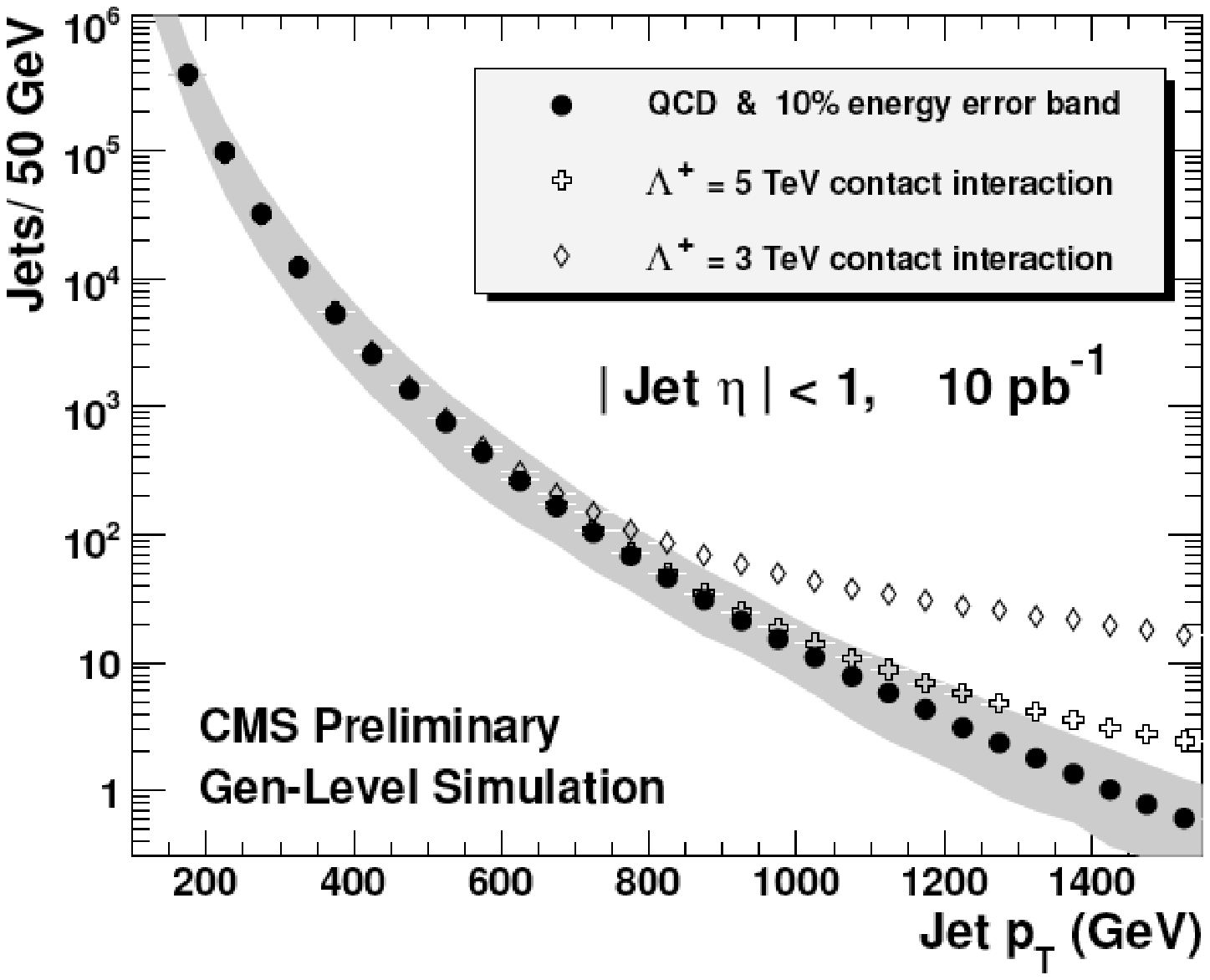}
    \includegraphics[height=4.6cm]{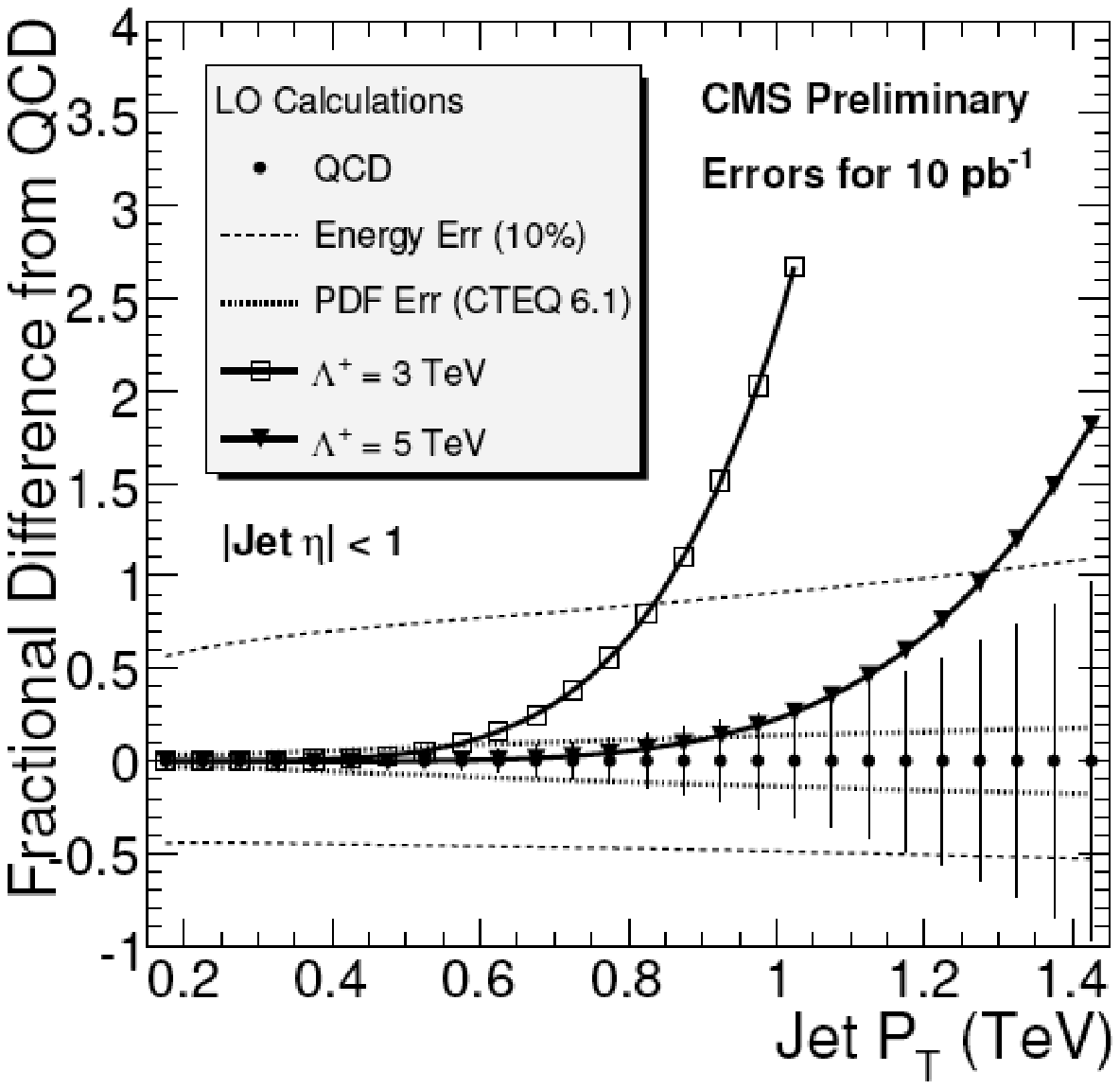}
    \caption{Left: inclusive jet cross-section in the central rapidity ($|\eta|<1$) with $10\,pb^{-1}$ at $\sqrt{s}=14\,TeV$ p-p collisions. The shaded band is an estimate of the dominant experimental uncertainties. Right: deviation from QCD due to contact interaction, compared to systematic and theoretical uncertainties.}
    \label{fig:inclusive}
  \end{center}
\end{figure}

The measurement of the inclusive jet cross-section, as a function of the jet $p_T$ is critical
for the commissioning of the jet object and the understanding of the detectors. Moreover, it is sensitive to new physics,
such as contact interactions and quark compositeness. Even with large experimental systematic uncertainties (dominated
by the JES and followed by the luminosity uncertainty), physics beyond the Standard Model will manifest itself as large deviation from
QCD at high jet $p_T$. With $10\,pb^{-1}$ at $\sqrt{s}=14\,TeV$ p-p collisions, a contact interaction at energy scale $\Lambda=3\,TeV$ can be clearly seen
(the Tevatron excluded limit is $\Lambda<2.7\,TeV$~\cite{bib:D0_CONTACT}). In Fig.~\ref{fig:inclusive} the contact interaction signal is compared to the QCD prediction and the dominant systematic uncertainties.
It should be noted though that in the absence of new physics signal, the inclusive jet cross-section cannot be used as a precision QCD measurement due to the large experimental uncertainties. This can only be achieved if the JES uncertainty is $\sim 1-2\%$.

\section{Dijet Mass and Dijet Ratio}
\begin{figure}[htbp]
  \begin{center}
     \includegraphics[height=4.6cm]{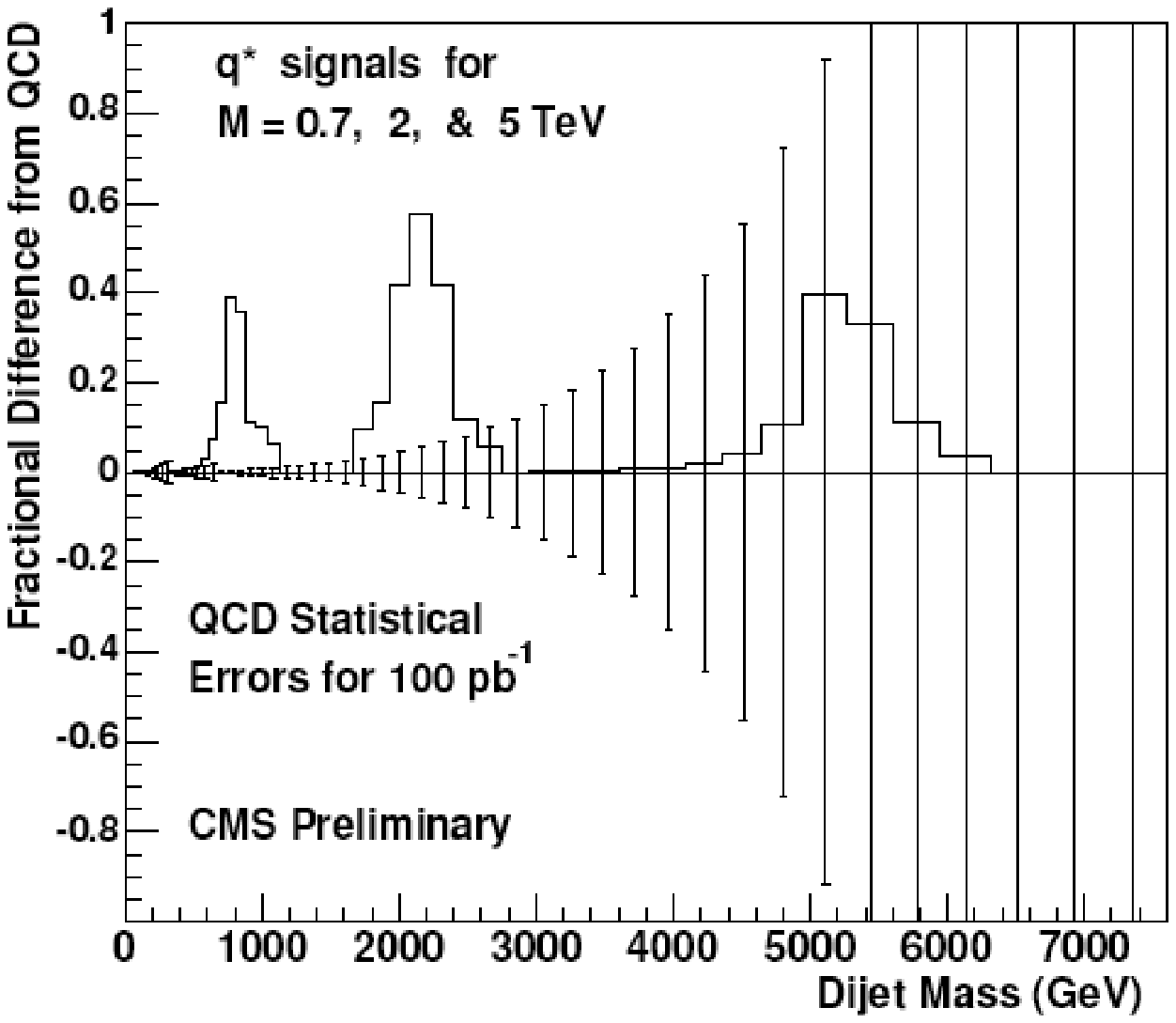}
    \includegraphics[height=4.6cm]{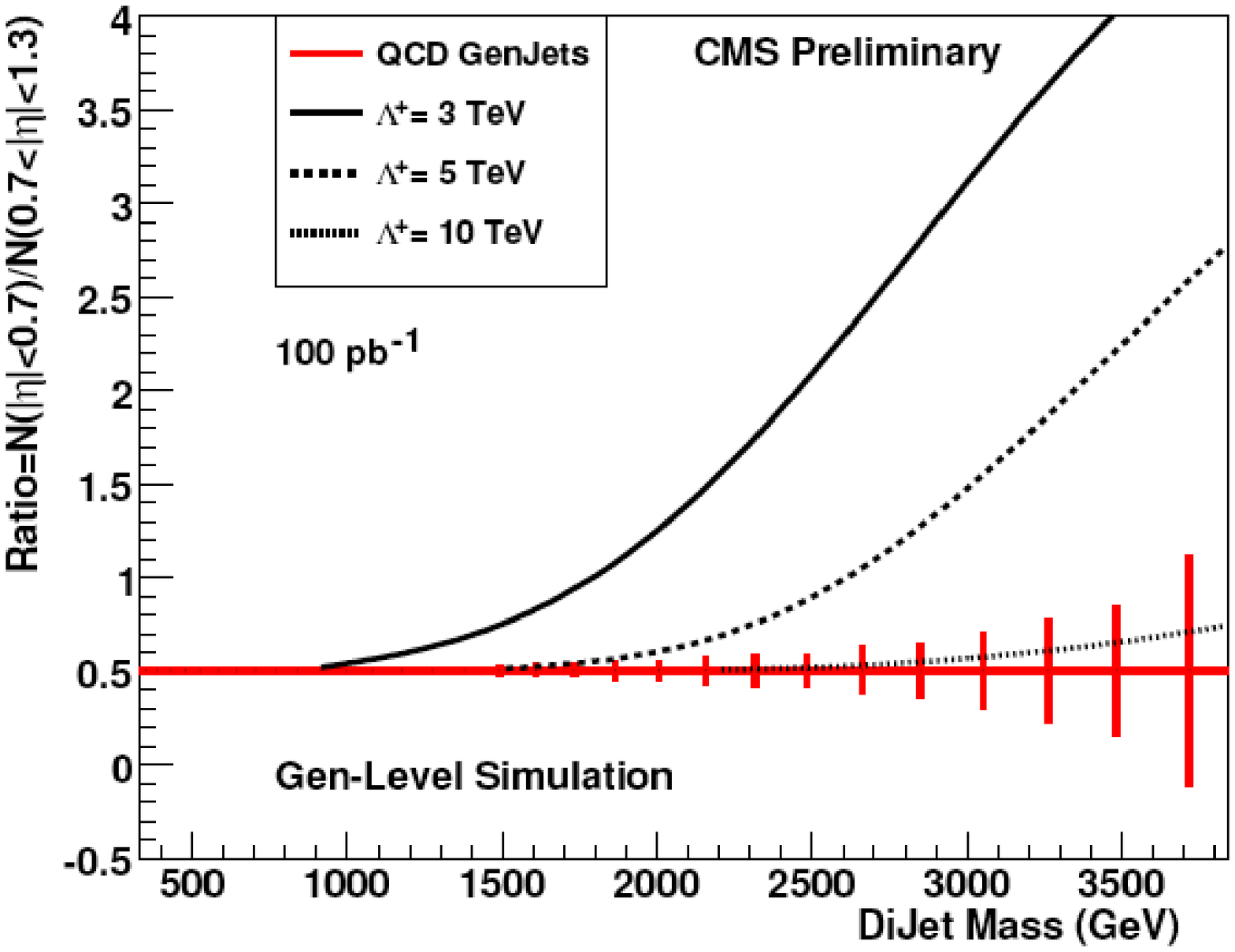}
    \caption{Left: fractional deviation of excited quark production from QCD, as a function of the dijet mass, for various resonance masses. 
    The errors correspond to the statistical uncertainty expected with $100\,pb^{-1}$ at $\sqrt{s}=14\,TeV$ pp collisions. Right: the dijet ratio
    observable for QCD (flat line) compared to contact interaction.}
    \label{fig:mass}
  \end{center}
\end{figure}

Another observable of interest is the dijet production cross-section, as a function
of the invariant mass of the two jets with the highest $p_T$ in an event. This measurement
will be used to confront the QCD predictions at transverse momentum scales, far beyond
any previous experiment. It can also be used to search for new physics~\cite{bib:CMS_DIJETS}, such as resonances 
(e.g. excited quark) that decay to two jets. In Fig.\ref{fig:mass}(left) the fractional deviation of excited quark signal from QCD
is compared to the statistical uncertainty corresponding to $100\,pb^{-1}$ of data with pp collisions at $\sqrt{s}=14\,TeV$.  
However, the direct measurement of the dijet cross-section is dominated by the JES systematic uncertainty and  cannot
be used as a precision QCD test with early LHC data.

Observables sensitive to the angular properties of the dijet production can be used early on to detect
deviations from QCD predictions. While the QCD production is dominated by t-channel scattering, new physics
signals tend to be more isotropic (s-channel). When enough data will be available and the detectors are understood, the
study of the angular distributions will provide maximum sensitivity to new physics. During the early data taking however, the
study of the dijet ratio as a function of the dijet mass will be a robust measure of the angular production properties.
The dijet ratio is defined as $R=N(|\eta|<0.7)/N(0.7<|\eta|<1.3)$ where $N(|\eta|<0.7)$ is the number of dijet events with both
jets observed in the central rapidity region and $N(0.7<|\eta|<1.3)$ is the number of events with both jets observed in the region
$0.7<|\eta|<1.3$. In Fig.\ref{fig:mass}(right) the QCD prediction for the dijet ratio is approximately flat at the value of $\sim 0.5$
while new physics, such as contact interaction, leads to large deviation with increasing dijet mass. In the dijet ratio measurement
important experimental systematic uncertainties (JES, luminosity) cancel and therefore this measurement is suitable for early data.

\section{Other QCD studies}
\begin{figure}[htbp]
  \begin{center}
    \includegraphics[height=4.7cm]{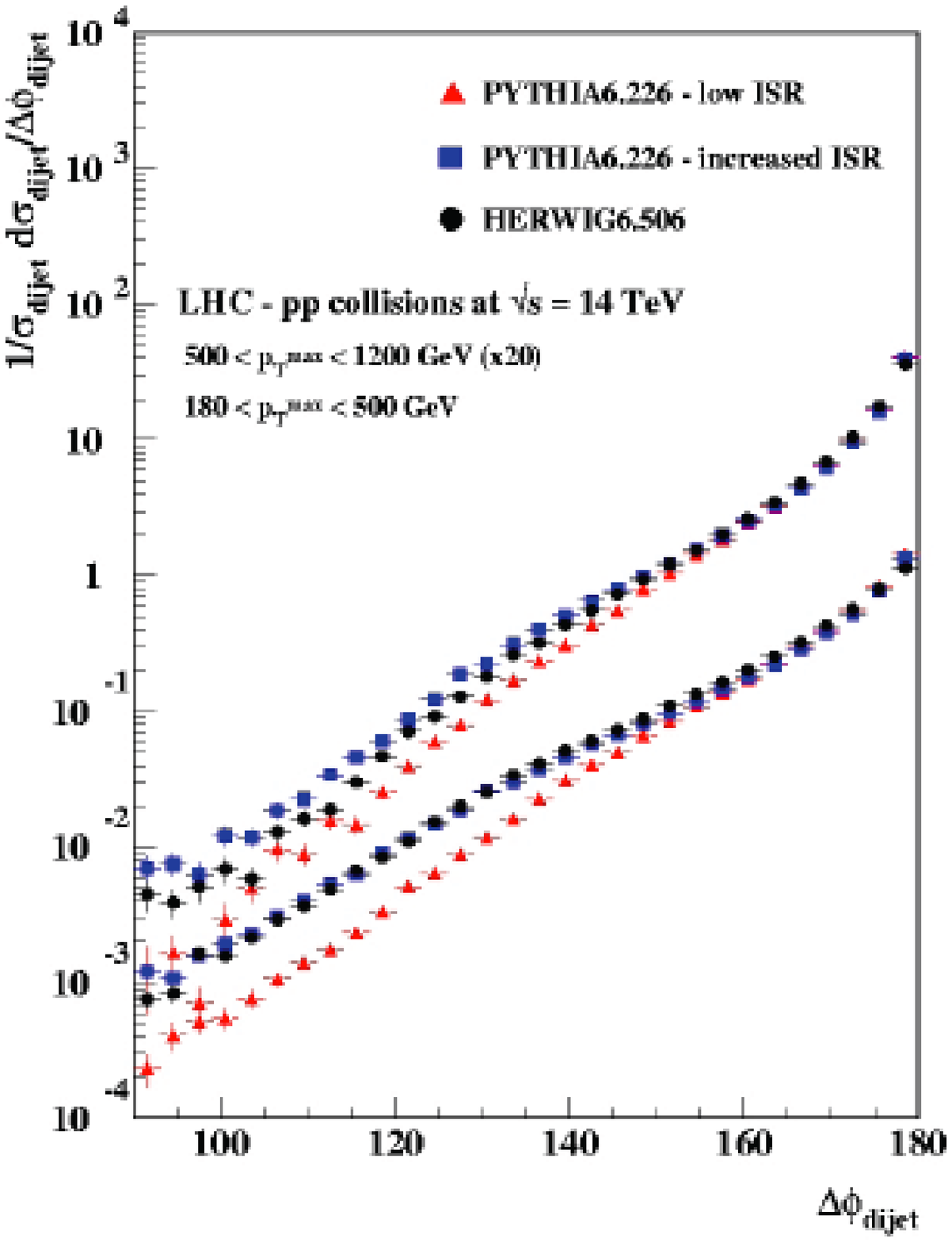}
    \includegraphics[height=4.7cm]{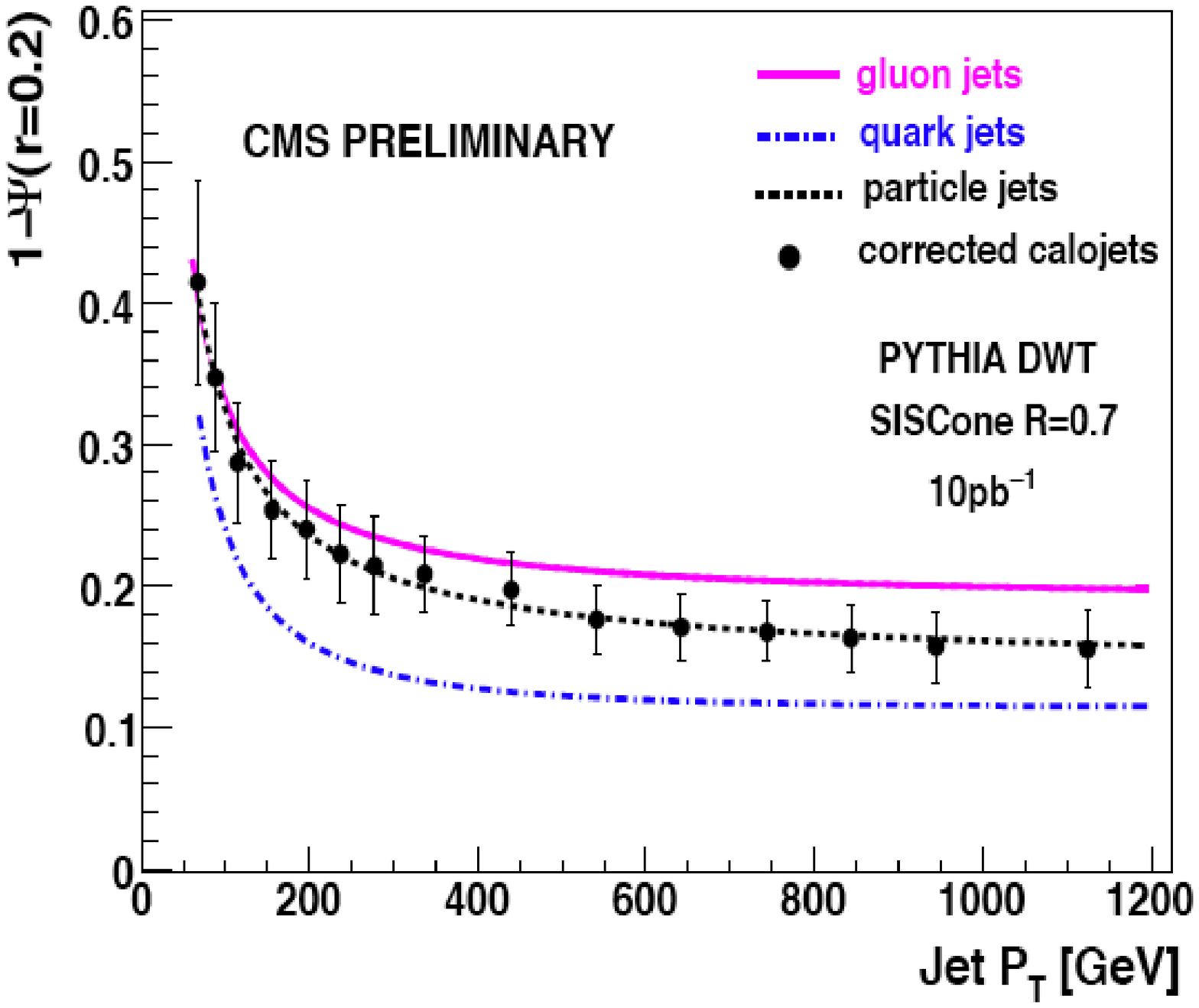}
    \caption{Left: dijet azimuthal decorrelation as predicted by different generators from ATLAS simulation. Right: jet shapes variable as a function of the 
    jet $p_T$ from CMS simulation at $\sqrt{s}=14\,TeV$ pp collisions. The error bars reflect the combined statistical and systematic uncertainty.}
    \label{fig:jetShapes}
  \end{center}
\end{figure}

In addition to the jet measurements which are sensitive to new physics, dedicated QCD studies are also feasible with early data. The dijet
azimuthal decorrelation observable ($1/N dN/d\Delta\phi$) is sensitive to the underlying QCD dynamics~\cite{bib:AZIMUTHAL} and at the same time unaffected by the
JES uncertainty. In Fig.\ref{fig:jetShapes} (left) typical dijet azimuthal decorrelation distributions are shown from different Monte Carlo generators
(ATLAS simulation). The shape at low $\Delta\phi$ values is sensitive to the fragmentation and hadronization modelling and this measurement can be done with early data in order to confront the QCD predictions and tune the Monte Carlo event generators.

The jet structure is also sensitive to the showering and fragmentation modelling in the Monte Carlo generators. It can be quantified by studying the 
transverse momentum distribution of the jet constituents~\cite{bib:CMS_JET_SHAPES}. The measurement is moderately affected by the JES uncertainty
and can be performed with early data. In Fig.\ref{fig:jetShapes}(right), the jet shape variable is shown as a function of the jet $p_T$.
At low $p_T$ the QCD jets originate mostly from gluons (wider calorimeter shower) while at higher $p_T$ the quark fraction increases and the jets become narrower.
 
\section{Conclusions}
Jet measurements at LHC will be used to confront the predictions of QCD while at the same time being sensitive to new physics signals.
Despite the large systematic uncertainties expected at startup, mainly due to the JES, the unprecedented $p_T$ reach with pp collisions
at $14\,TeV$ will allow the exploration of the $TeV$ scale. Precision QCD measurements however will not be feasible until the experimental
uncertainties are sufficiently understood and reduced.
Both CMS and ATLAS have a rich QCD physics program with jets which will be employed as soon as collision data become available.

\end{document}